\documentclass{appolb}
\usepackage[utf8]{inputenc}
\usepackage{epsf}
\usepackage{latexsym,amssymb,euscript}
\usepackage[dvips]{graphicx}
\usepackage[numbers,sort&compress]{natbib}
\usepackage{amsmath}
\usepackage{nicefrac}
\usepackage{slashed}
\usepackage{booktabs}
\usepackage{hyperref}
\usepackage{braket}
\usepackage{chngcntr}
\usepackage{bbold}
\usepackage{graphics}
\usepackage{graphicx}
\usepackage{mciteplus}
\usepackage{caption}
\usepackage{subcaption}
\usepackage{pdfpages}
\usepackage[titletoc]{appendix}
\graphicspath{{./figures/}}
\hypersetup{
 linktocpage = true,
 urlcolor = black,
 colorlinks = true,
 linkcolor = urlblue,
 anchorcolor = urlblue,
 citecolor = urlblue,
 pdfstartview = {XYZ null null 1.25} 
           }
\usepackage{pstricks}
\usepackage{color}
\usepackage{xcolor}
\definecolor{urlblue}{rgb}{0.2,0.4,0.7}
\definecolor{citegreen}{rgb}{0,0.4,0.2}
\definecolor{linkred}{rgb}{0.9,0.2,0.1}
\usepackage{float}
\definecolor{orcidlogocol}{HTML}{A6CE39}
\usepackage{fancyhdr}
\usepackage[utf8]{inputenc}
\usepackage[normalem]{ulem}
\usepackage{mathtools}
\usepackage{setspace}

\newcommand{{\HFNRevo}}{\tt HF-NRevo}

\begin{document}
\title{Heavy-Flavor Fragmentation: \\ The QCD Portal to Exotic Matter%
\thanks{Presented at the ``Excited QCD 2026'' Workshop, Universidad de Granada, Carmen de la Victoria (Spain), January 8-14, 2026.}
}
\author{Francesco Giovanni Celiberto
\address{Universidad de Alcal\'a (UAH), E-28805 Alcal\'a de Henares, Madrid, Spain}
}
\maketitle

\begin{abstract}
We investigate the core dynamics behind exotic matter formation via the {\tt TQ4Q1.1} set of collinear fragmentation functions for fully charmed or bottomed tetraquarks in three quantum configurations: scalar ($0^{++}$), axial vector ($1^{+-}$), and tensor ($2^{++}$). 
We adopt leading-power single-parton fragmentation within a nonrelativistic QCD framework tailored to tetraquark Fock states.
Initial-scale inputs are constructed from updated gluon- and heavy-quark channels, and evolved through threshold-consistent DGLAP within HF-NRevo.
We present the first systematic propagation of uncertainties from color-composite long-distance matrix elements governing tetraquark hadronization.
This study advances the connection between hadronic structure, precision QCD, and exotic matter.
\end{abstract}

\vspace{-0.25cm}

\section{Introduction}
\label{sec:introduction}

Hadrons containing heavy quarks play a central role in probing the fundamental dynamics of the strong interaction.
Their large masses, lying well above the nonperturbative QCD scale, enable a controlled separation between perturbative and nonperturbative effects, making them ideal systems for precision studies.
At the same time, their possible couplings to physics beyond the Standard Model render them valuable tools in the broader search for New Physics.
A key open question in hadron physics concerns how QCD organizes matter beyond the conventional quark–antiquark and three-quark configurations.
Exotic states such as tetraquarks and pentaquarks, characterized by a genuine multiquark structure, challenge our understanding of color confinement and hadronization mechanisms, due to their extended valence content and internal complexity.
Among them, fully heavy tetraquarks provide a particularly clean laboratory, where the interplay between perturbative production and nonperturbative binding can be investigated in a theoretically controlled setting.
Theoretical descriptions of heavy-hadron formation remain challenging.
While significant progress has been achieved for quarkonia (often regarded as the ``hydrogen atom'' of QCD) no framework fully captures all aspects of hadronization across different kinematic regimes.
Non-Relativistic QCD (NRQCD)~\cite{Caswell:1985ui,Bodwin:1994jh} provides a systematic approach by organizing production processes in terms of perturbative Short-Distance Coefficients (SDCs) and nonperturbative Long-Distance Matrix Elements (LDMEs), summed over all possible Fock states.
Within this framework, quarkonium production can be described both at low transverse momenta, dominated by short-distance heavy-quark pair creation, and at moderate-to-high transverse masses, where single-parton fragmentation becomes the leading mechanism~\cite{Cacciari:1994dr}.
Early studies of fragmentation into quarkonia, starting from leading-order calculations for gluon and heavy-quark channels~\cite{Braaten:1993rw,Braaten:1993mp} and later improved at NLO~\cite{Zheng:2019gnb,Zheng:2021sdo_short}, enabled the construction of the first VFNS-consistent FF sets~\cite{Celiberto:2022dyf_short,Celiberto:2023fzz}, including extensions to $B_c$ mesons~\cite{Celiberto:2022keu,Celiberto:2024omj}.
These developments, supported by comparisons with LHC data, established the validity of fragmentation-based descriptions in the high-energy regime and motivated the formulation of a consistent evolution framework.
Building on this progress, the NRQCD-based description of fragmentation has been extended to exotic multiquark systems.
Experimental observations of double-$J/\psi$ final states~\cite{LHCb:2020bwg,ATLAS:2023bft_short,CMS:2023owd_short} have been interpreted as signatures of compact tetraquark configurations~\cite{Zhang:2020hoh_short,Zhu:2020xni}, where two heavy-quark pairs are produced at short distances and subsequently bind into a multiquark state.
Initial NRQCD inputs for gluon fragmentation into fully heavy tetraquarks were derived in~\cite{Feng:2020riv}, paving the way for the first VFNS-consistent FF constructions for exotic hadrons.
This program led to the development of the {\tt TQHL1.0} sets for heavy-light tetraquarks~\cite{Celiberto:2023rzw_short,Celiberto:2024mrq}, followed by the {\tt TQ4Q1.x} and {\tt TQHL1.1} releases~\cite{Celiberto:2024mab_short,Celiberto:2025dfe,Celiberto:2025ziy,Celiberto:2024beg_short}, which incorporate NRQCD-based inputs for both gluon and heavy-quark channels and extend the description to fully heavy configurations.
These developments have established a unified strategy for describing exotic-hadron production within QCD, further extended to systems such as fully heavy pentaquarks and triply heavy baryons~\cite{Celiberto:2025ipt,Celiberto:2025ogy}.
Recent studies have also explored tetraquark production in forward kinematics~\cite{Celiberto:2025vra}, where the interplay between gluon-initiated and heavy-quark-initiated channels reveals distinct sensitivities to the underlying partonic structure, including possible intrinsic-charm contributions.

In this work, we focus on collinear fragmentation into fully heavy tetra\-quarks using the {\tt TQ4Q1.1} FF sets, constructed within the HF-NRevo evolution scheme~\cite{Celiberto:2025euy_short,Celiberto:2024mex_article,Celiberto:2024bxu,Celiberto:2024rxa,Celiberto:2025xvy_short,Celiberto:2026rzi_short}.
This approach consistently combines NRQCD initial-scale inputs at NLO, threshold-aware DGLAP evolution, and a systematic treatment of Missing Higher-Order Uncertainties (MHOUs) through Monte-Carlo-like methods~\cite{Forte:2002fg}, providing a predictive and uncertainty-controlled description of exotic multiquark fragmentation.

\section{Heavy-Tetraquark Fragmentation from HF-NRevo}
\label{sec:HFNrevo}

\begin{figure*}[!t]
\centering

   \includegraphics[scale=0.315,clip]{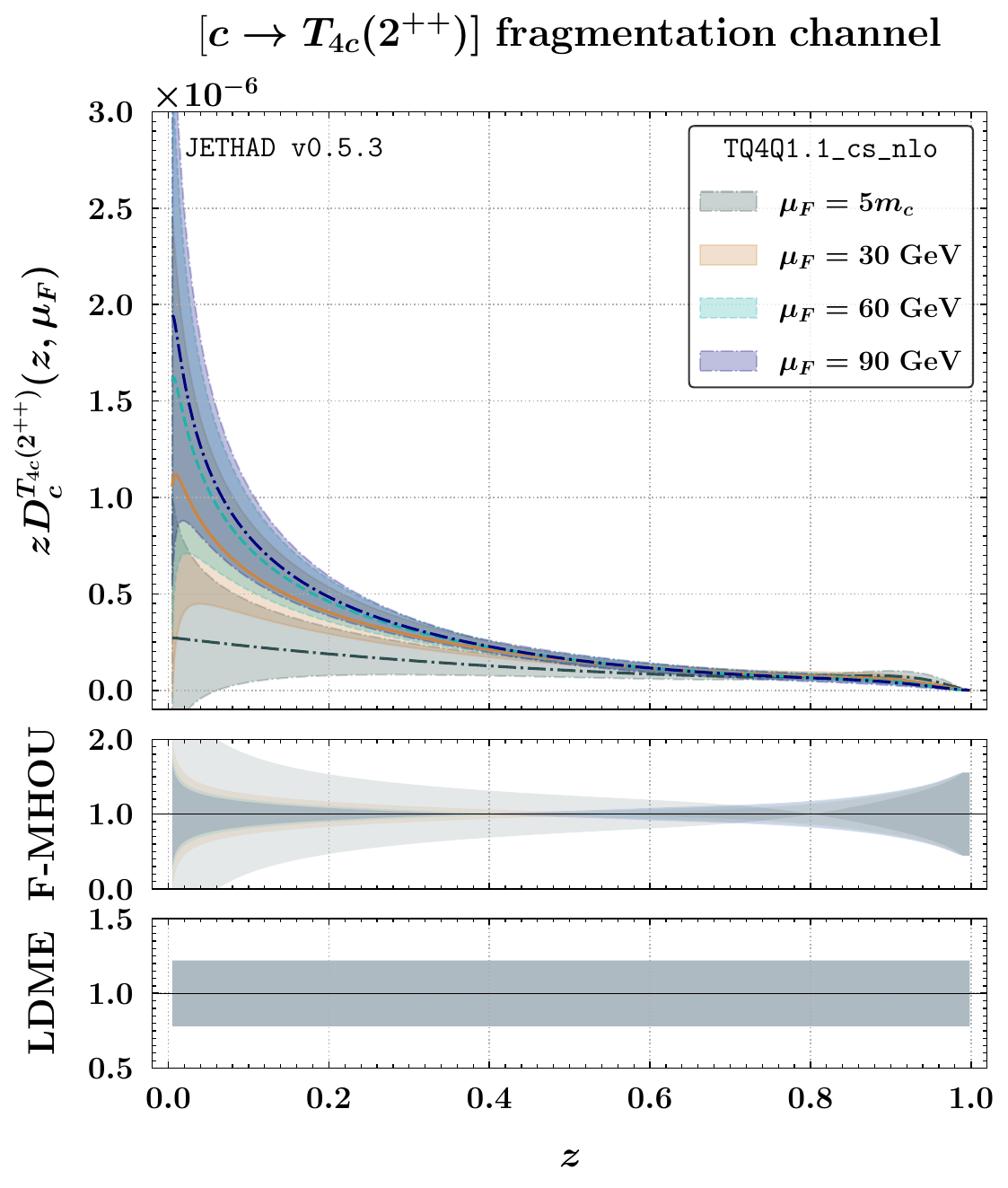}
   \hspace{0.20cm}
   \includegraphics[scale=0.315,clip]{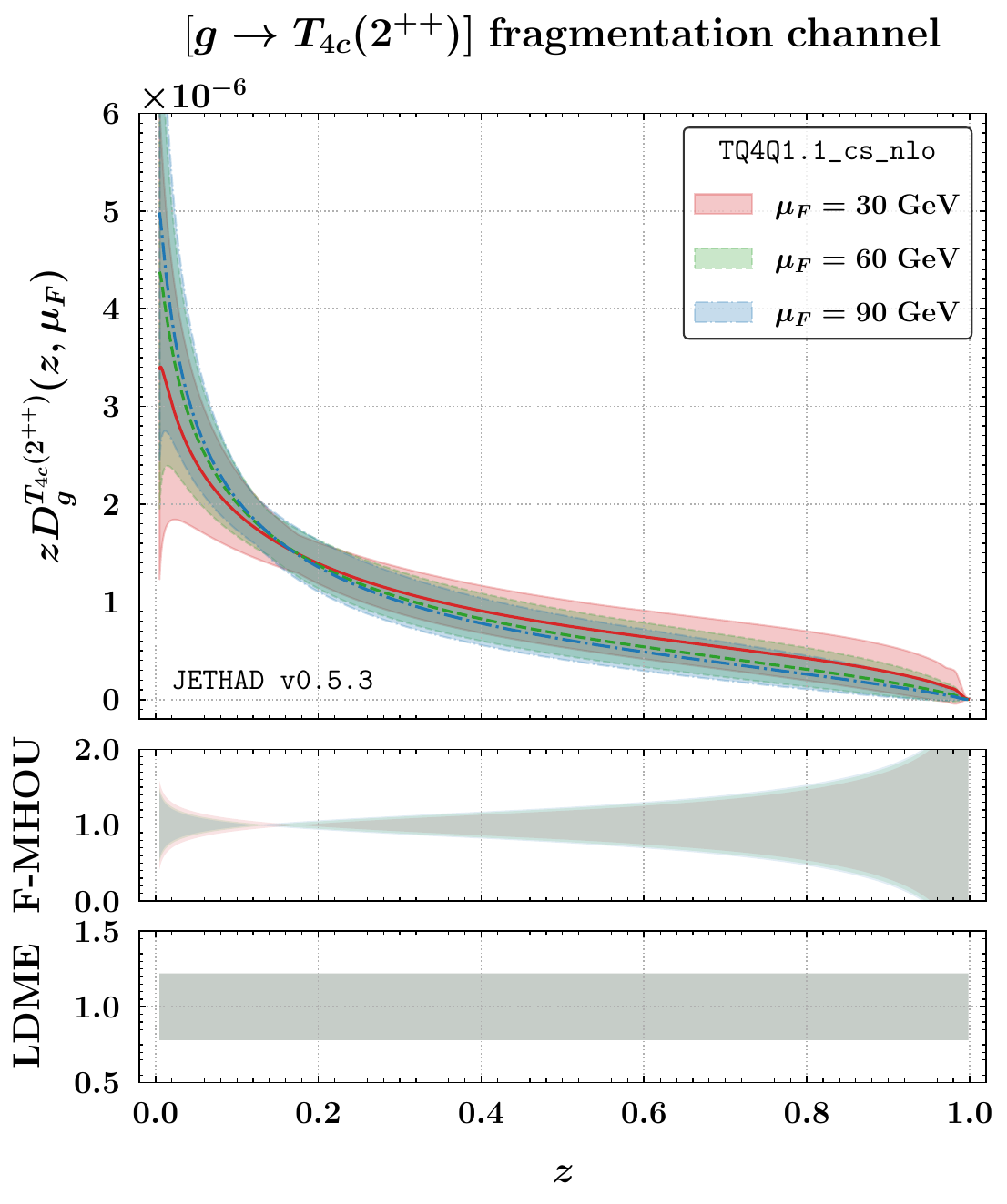}

\caption{$z$-shape of the charm (left) and gluon (right) {\tt TQ4Q1.1} FFs for tensor tetraquarks $T_{4c}(2^{++})$ (left) at various energy scales.
Main-panel bands include both F-MHOU and LDME uncertainties, while the lower panels separately show F-MHOU replica envelopes and LDME variations as ratios to the central prediction.
\vspace{-0.40cm}
}

\label{fig:FFs}
\end{figure*}

We present the DGLAP evolution of NRQCD initial-scale inputs leading to the {\tt TQ4Q1.1} collinear FFs for fully heavy tetraquarks.
In contrast to light-hadron fragmentation, both gluon and heavy-quark channels are characterized by distinct evolution thresholds, fixed by the kinematics of the underlying perturbative splittings $[g \to (Q\bar{Q}Q\bar{Q})]$ and $[Q,\bar{Q} \to (Q\bar{Q}Q\bar{Q})+Q,\bar{Q}]$.
These mechanisms naturally set $\mu_{F,0}(g \to T_{4c})=4m_Q$ and $\mu_{F,0}(Q \to T_{4c})=5m_Q$ as the initial scales for gluon and (anti)quark fragmentation, respectively, introducing a genuine multi-threshold structure already at the perturbative level.
To consistently account for this structure, we adopt the HF-NRevo scheme~\cite{Celiberto:2025euy_short,Celiberto:2024mex_article,Celiberto:2024bxu,Celiberto:2024rxa,Celiberto:2025xvy_short,Celiberto:2026rzi_short}, a framework designed for the evolution of heavy-hadron FFs with nonrelativistic inputs.
Building on the same conceptual pillars developed in quarkonium studies, namely interpretation, evolution, and uncertainties, HF-NRevo enables a coherent connection between low and high transverse-momentum regimes while preserving the physical meaning of threshold scales.
In the tetraquark case, this interpretation is further enriched by the composite color structure of the final state, encoded in NRQCD LDMEs, which act as genuinely nonperturbative inputs coupled to perturbative short-distance coefficients.
The evolution proceeds in two stages.
First, the gluon FF, initialized at $\mu_{F,0}=4m_Q$, evolves up to $\mu_{F,0}=5m_Q$ through a single-channel dynamics driven by the LO $P_{gg}$ kernel, generating only collinear gluon radiation.
This step, involving a decoupled evolution sector, is handled analytically via the \textsc{symJethad} plugin~\cite{Celiberto:2020wpk,Celiberto:2022rfj,Celiberto:2023fzz,Celiberto:2024mrq,Celiberto:2024swu,Celiberto:2025_P5Q_review,Celiberto:2025csa,Celiberto:2026zed}, which allows for exact control of threshold effects at the symbolic level (for  applications on  precision QCD phenomenology and the hadron structure at low-$x$ with \textsc{(sym)Jethad}, see Refs.~\cite{Celiberto:2017ius_short,Celiberto:2015yba,Celiberto:2016ygs,Celiberto:2022gji,Celiberto:2016hae,Celiberto:2017ptm,Bolognino:2018oth,Bolognino:2019cac,Celiberto:2020rxb_short,Celiberto:2022kxx,Celiberto:2020tmb,Celiberto:2023rtu,Celiberto:2023uuk,Celiberto:2023eba,Celiberto:2023nym,Celiberto:2023rqp,Celiberto:2024mdt,Celiberto:2024bfu,Celiberto:2025edg,Bolognino:2021mrc,Celiberto:2017nyx,Bolognino:2019ouc,Bolognino:2019yls,Celiberto:2021dzy,Celiberto:2021fdp,Celiberto:2022grc} and~\cite{Bolognino:2018rhb,Bolognino:2018mlw,Bolognino:2019pba,Celiberto:2019slj,Bolognino:2021niq,Bolognino:2021gjm,Celiberto:2018muu}, respectively).
At the heavy-quark threshold ($Q_0=5m_Q$), the evolved gluon FF is consistently matched to the NRQCD input of the (anti)quark channel.
From this ``evolution-ready'' configuration, a full NLO DGLAP evolution including all active parton species is performed numerically, yielding the final {\tt TQ4Q1.1} sets.
Uncertainties associated with missing higher-order contributions in the fragmentation sector (F-MHOUs) are estimated through variations of the initial evolution scale $Q_0$ by a factor of two around its central value.
Each variation defines a replica that is independently evolved, producing an uncertainty envelope that quantifies the sensitivity of the FFs to threshold choices.
This procedure closely parallels modern PDF analyses based on theory-covariance matrices~\cite{NNPDF:2024dpb} or Monte Carlo scale variations~\cite{Kassabov:2022orn}, and allows for a consistent propagation of uncertainties to collider observables.
These perturbative uncertainties complement those arising from the nonperturbative LDMEs, which primarily affect the overall normalization of the FFs.
Light-quark and nonconstituent heavy-quark channels are not included in this release, as their contributions are suppressed by one or more orders of magnitude relative to gluon fragmentation.
For simplicity, we present here two fragmentation channels.
Left and right plots of Fig.~\ref{fig:FFs} respectively show $[c \to T_{4c}(2^{++})]$ and $[g \to T_{4c}(2^{++})]$ {\tt TQ4Q1.1} NLO FFs for $\mu_F$ values ranging from 30 to 90~GeV.

\section{Towards Multimodal Fragmentation}
\label{sec:conclusions}

sing HF-NRevo, we developed the {\tt TQ4Q1.1} collinear FF sets for fully heavy tetraquarks, based on NRQCD inputs for gluon and constituent-heavy-quark channels.
The framework consistently incorporates the multi-threshold structure of fragmentation and enables a unified DGLAP evolution from the matching scale to collider energies.
Uncertainties are quantified through a replica-like approach, combining F-MHOUs from threshold variations with nonperturbative LDME effects.
The {\tt TQ4Q1.1} functions provide the first robust and publicly available description of fully heavy tetraquark fragmentation, offering a baseline for studies of exotic production at current and future facilities.
They allow for a systematic exploration of the interplay between perturbative dynamics and nonperturbative formation, with applications at the LHC, its high-luminosity upgrade, and future colliders.
More generally, this work opens the way to a multimodal description of heavy-hadron fragmentation, where different production mechanisms (direct, diquark, and higher Fock states) are consistently combined.
This approach connects precision QCD with exotic-hadron phenomenology and provides a basis for future data-driven and machine-learning developments.

\section*{Acknowledgments}
\label{sec:acknowledgments}

This work is supported by the Atracci\'on de Talento Grant n. 2022-T1/TIC-24176 (Madrid, Spain).

\vspace{-0.05cm}
\begingroup
\setstretch{0.6}
\bibliographystyle{apsrev}
\bibliography{proc_references}

\end{document}